\documentclass{PoS}
\usepackage{amsmath}
\title{Multi GPU Performance of Conjugate Gradient Algorithm
with Staggered Fermions}

\ShortTitle{Multi GPU Performance}

\author{\speaker{Hyung-Jin Kim}, Weonjong Lee \\
  Lattice Gauge Theory Research Center, FPRD, and CTP \\
  Department of Physics and Astronomy, 
  Seoul National University, Seoul, 151-747, South Korea \\
  E-mail: \email{windy510@gmail.com}, \email{wlee@snu.ac.kr}}

\abstract{ We report results of the performance test of GPUs obtained
  using the conjugate gradient (CG) algorithm for staggered fermions
  on the MILC fine lattice ($28^3 \times 96$).
  We use GPUs of nVIDIA GTX 295 model for the test.
  When we turn off the MPI communication and use only a single GPU,
  the performance is 35 giga flops in double precision, which
  corresponds to 47\% of the peak. 
  When we turn on the MPI communication and use multi-GPUs,
  the performance is reduced down to 12.3 giga flops.
  The data transfer through the infiniband network and PCI-E bus I/O
  is a main bottle neck.
  We suggest two potential solutions of how to optimize the data
  transfer.
}

\FullConference{
  The XXVIII International Symposium on Lattice Field Theory, Lattice2010\\
  June 14-19, 2010\\
  Villasimius, Italy
}

\begin{document}

\section{Introduction}

There has been a significant progress in accelerating the
computational speed.
CPU has kept improving its computing power but does not yet quench the
thirst of those demanding users who need more computing power for
their numerical challenges such as lattice QCD.
One bailout was to make a CPU cluster to do the parallel processing,
which also causes a problem of how to handle the scalable
communication between nodes.
Lattice group at Columbia University developed customized machines for
lattice QCD such as QCDOC \cite{QCDOC}, which has a great performance
in the nearest neighbor communication.

At present, graphic processing units (GPU) opens a new era for high
performance computing.
GPU is originally designed to handle 3-dimensional graphic images.
GPU is composed of many tiny multi processors which are more
appropriate architecture to handle the single instruction
multiple data (SIMD) than multi cores of CPU.
In addition, the evolution of GPU is very fast. 
In Table \ref{tab:gpu-1}, we summarize the peak performance of the
current highend GPUs in the market \cite{nvidia_ref}.
The enormous computing power of GPUs is highly beneficial to lattice
QCD.
\begin{table}[h!]
\begin{center}
\begin{tabular}{c || c | c | c}
\hline
model & peak speed (SP) & peak speed (DP) & memory bandwidth (GB/s)\\
\hline
GTX 285 & 1062.72 GF & 88.6 GF &  159.0 \\
GTX 295 & 1788.48 GF & 149 GF  &  223.8 \\
GTX 480 & 1344.96 GF & 168 GF  &  177.4 \\
C2050   & 1.03 TF    & 515 GF  &  144.0 \\
\hline
\end{tabular}
\end{center}
\caption{GPU models and their peak performance.
  Here, SP (DP) means single (double) precision.
  GF (TF) means giga (tera) flops.
  GB/s means giga bytes per second.
}
\label{tab:gpu-1}
\end{table}

There are several ways to implement the GPU code in the market:
Nvidia CUDA API, Open Graphic Library (Open GL),and Open
Computing Language (Open CL).
In this paper, we focus on CUDA and its applications.
The CUDA provides us a user-friendly programming environment based on
the C, C++ programming language for GPU.
All of our code are compiled and tested in CUDA version 2.3 and
compute capability 1.3 mode.
We make the CUDA version of CG subroutine that is implemented as a
part of the Columbia Physics System (CPS) library.

Let us turn to the hardware. 
We have constructed a GPU cluster whose name is David.
The David cluster has 32 nodes which are connected through the 20
giga bit infiniband network.
Each node has a Intel core i7 920 Processor and two of nVIDIA GTX 295
graphic cards\footnote{ At present, GTX 295 graphic cards are
  upgraded to nVIDIA GTX 480.}.
We use LINUX Cent OS version 5.5 as an operating system of this cluster. 

\section{CG Implementation using CUDA}
Conjugate gradient (CG) algorithm \cite{CG_paper} is an iterative
method for solving a linear algebraic equation of the following form.
\begin{equation}
\mathbf{b} = A\mathbf{x}\,,
\end{equation}
where $A$ is a $n \times n$ positive definite Hermitian matrix. 
$\mathbf{x}$ and $\mathbf{b}$ are complex vectors in the n dimensional
space.
Matrix $A$ and vector $\mathbf{b}$ are given and $\mathbf{x}$ vector is
a solution that we want to obtain.
Using the CG method we can get the solution $\mathbf{x}$ up to the
numerical precision that we want to achieve.
\begin{figure}[h]
\begin{center}
\framebox{\parbox{10cm}{\begin{quote}
$\mathbf{r} = \mathbf{b} - A\mathbf{x} \mbox{\qquad\qquad\qquad} \mathbf{r} \mbox{: residual vector}$\\
$\mathbf{d} = \mathbf{r} \mbox{\qquad\qquad\qquad\qquad} \mathbf{d} \mbox{: directional vector}$\\
$\delta_{new} = \mathbf{r}^{\dagger}\mathbf{r}\mbox{\qquad\qquad\qquad} \epsilon \mbox{: tolerance}$\\
$\delta_{0} = \delta_{new}$\\
for$(i=0; i<N_{dim} \& \delta_{new} > \epsilon^2\delta_{0}; ++i)\{$\\
\qquad$\alpha = \delta_{new}/\mathbf{d}^{\dagger}A\mathbf{d}$\\
\qquad$\mathbf{x} = \mathbf{x} + \alpha \mathbf{d}$\\
\qquad$\mathbf{r} = \mathbf{r} - \alpha A\mathbf{d}$\\
\qquad$\delta_{old} = \delta_{new}$\\
\qquad$\delta_{new} = \mathbf{r}^{\dagger}\mathbf{r}$
\qquad$\beta = \delta_{new}/\delta_{old}$\\
\qquad$\mathbf{d} = \mathbf{r} + \beta \mathbf{d}$\quad$\}$
\end{quote}}}
\end{center}
\caption{Conjugate gradient
  algorithm\label{CG_structure}}
\end{figure}
In Fig.~\ref{CG_structure}, we show the structure of CG algorithm.
In the CG sequence, we have a number of linear algebra equations
such as vector addition, dot product, and scalar multiplication
and so on.
All of these linear operations are implemented using CUBLAS
library \cite{nvidia_CUBLAS}, except for the Dirac operation.
Note that these operations are not dominant in CG operation, and so
CUBLAS Library is good enough to handle them.
In Fig.~\ref{CG_structure}, $A\mathbf{d}$ and $A\mathbf{x}$ are Dirac
operations with staggered fermions \cite{stagger}. Basically, the
Dirac operation is a matrix-vector multiplication.
This is dominant in CG.
The matrix $A$ is defined as follows.
\begin{equation}
\mathbf{h} = A\mathbf{\chi}
\end{equation}
\begin{equation}
A = -D^{2} + m^{2} \quad \mbox{(m is quark mass)}
\end{equation}
\begin{equation}
D_{x,y} = U_{\mu}(x)\delta_{y,x+\mu} 
- U^{\dagger}_{\mu}(x-\mu)\delta_{y,x-\mu}
\end{equation}
\begin{equation}
D\chi(x) = \sum_{\mu}U_{\mu}(x)\chi(x+\mu) 
- U^{\dagger}_{\mu}(x-\mu)\chi(x-\mu)
\end{equation}
Here, note that the phase factor $\eta_\mu(x)$ is already multiplied
to the gauge link $U_\mu$ in the preconditioning part.
$h$ is a given source vector and $\chi$ is a staggered fermion field
which corresponds to the solution.
At a single site on the lattice, a single Dirac operation
of $D\chi(x)$ takes 1584 bytes of data transfer and 576 floating 
point calculations.
Let us consider a MILC fine lattice of $28^3 \times 96$.
A single Dirac operation $D\chi(x)$ over the entire even sites of the
lattice takes 0.6 billion floating point calculations and 1.6 giga
bytes of data transfer.
%-----------------------------------
% 0.6 giga / 75 giga  = 0.008 sec
% 1.6 giga / 100 giga = 0.016 sec
%----------------------------------
When we use GTX 295 GPUs, it is easy to find out that the bottle neck
is in the data transfer.

\section{CUDA CG code Optimization}
Starting CUDA programming itself is easy and straight-forward. 
We change 4 dimensional for-loops into the CUDA thread index.
The CUDA makes each core of the GPU perform the CG calculation
at multiple lattice sites in parallel.
This simple modification is easy but does not provide a good
performance. 
It runs at about 1 giga flops per GPU (GTX 295). 
This is only twice faster than CPU. 
Hence, we must improve the performance of the CG code, which 
we will explain in the following subsections in detail.

\subsection{Coalesced Memory Access}
\begin{figure}
\begin{center}
\includegraphics[width=12cm]{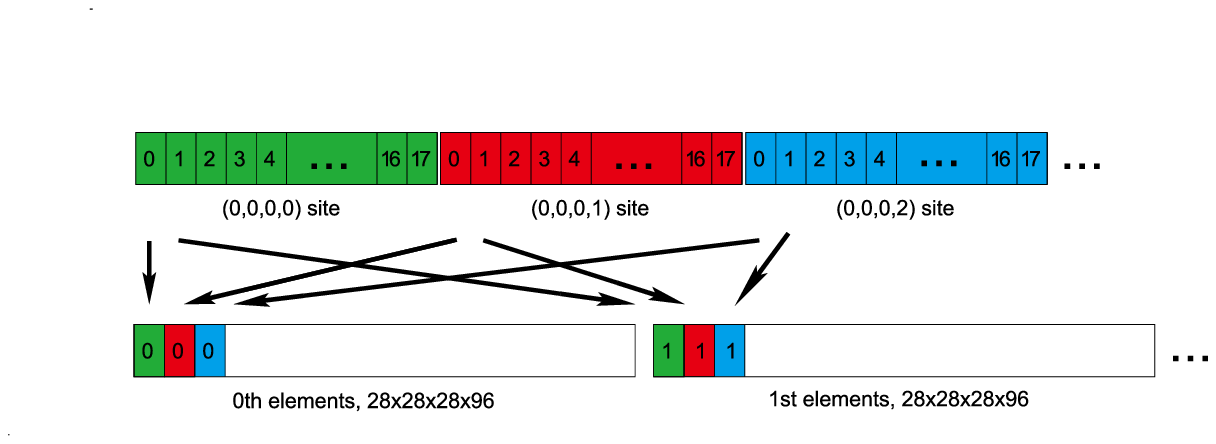}
\caption{Data structure for coalesced memory access.  CPU read gauge
  link data in serial way (upper part). But in GPU, multiple cores can
  read different data simultaneously. For maximum GPU memory
  bandwidth, the data addresses for simultaneous reading and writing
  should be continuous and multiples of 16 (lower part).}
\label{coal}
\end{center}
\end{figure}
Fig.~\ref{coal} shows the difference in the memory access pattern 
between CPU and GPU.
The CPU code accesses each memory address in sequential way.
In GPU, multiple cores access different memory addresses in parallel. 
If we do not align the memory address in a correct way, then we can
not get the maximum bandwidth of data transfer in GPU.
The correct way is called a coalesced memory access.
In the coalesced memory access, the data should be grouped into
a multiple of 16 as one can see in Fig.~\ref{coal}.
Then the data can be accessed simultaneously from multiple
cores at the highest speed \cite{nvidia_ref}.
We have changed our gauge link data and fermion vectors into
the coalesced access pattern.
As a result, we gain a 8.5 times performance enhancement.

\subsection{Register memory}
The GPUs of Nvidia have various kinds of memory area: 
constant memory, registers, shared memory, and so on.
Registers and shared memory are fast on-chip memory.
Hence, if we use them properly, we can reduce time for global memory
access.
As a result, we can enhance the overall performance of GPU.
In the Compute Capability 1.3\footnote{one of the runtime version in
  CUDA program}, each GPU (GTX 295) has 16k bytes of shared memory and
64k bytes of registers.
However, when we use double precision in floating point calculations,
shared memory has intrinsic memory bank-conflict problem\footnote{From
  compute capability 2.0, there is no bank-conflict using double
  precision in shared memory}.
The bank-conflict problem means that the data transfer between GPU and
shared memory becomes serialized and slows down by factor of 16
\cite{nvidia_ref}.
Hence, when we calculate in double precision, the register memory
shows significantly better performance.
Therefore if we use registers instead of shared memory, we can
speed up the program by factor of 3.
Hence, the performance reaches 25 giga flops.

\subsection{SU(3) reconstruction}

In the CG program, the bottle neck is in the data transfer between GPU
to GPU memory.
Hence, if we reduce the size of data transfer, then the performance
can be improved in principle.
Because gauge links are SU(3) matrices, we can use the 8 or 12
parameters for the SU(3) matrix reconstruction \cite{mixed_CUDA}. 
For example, the 3rd row of matrix elements can be reconstructed by
using the following equation.
\begin{equation}
\mathbf{c} = (\mathbf{a}\times\mathbf{b})^*
\label{SU3rec}
\end{equation}
Vectors $\mathbf{a,b,c}$ correspond 1st, 2nd ,3rd rows of a 
SU(3) matrix, respectively.
This reduces the amount of data transfer by 1/3. On the other hand,
we need to do additional 48 floating point calculations are needed so
that the floating point calculation increases by 67\%.
Furthermore, this reconstruction uses more registers.
As a result, this makes the CUDA occupancy\footnote{The CUDA occupancy
  is the ratio of the number of active threads involved in the
  calculation to the maximum number of threads.} decrease.
The net gain is about $-$7\%. 
Hence, the reconstruction method is not helpful.

\subsection{Additional Tune-up}
The CUDA occupancy is a good criterion to determine how efficiently
the kernel runs on the GPU.
It is not a simple object involved in many factors such as the number
of registers in use, size of shared memory in use, number of active
threads, and so on.
Hence, it is so complicated that there is no cure-all solution in
general.
\begin{equation}
\begin{pmatrix}
	x^{'}_1\\ x^{'}_2\\ x^{'}_3
\end{pmatrix}
=
\begin{pmatrix}
	a_1 & a_2 & a_3 \\ b_1 & b_2 & b_3 \\ c_1 & c_2 & c_3  \notag
\end{pmatrix}
\begin{pmatrix}
	x_1\\ x_2 \\ x_3
\end{pmatrix}
\end{equation}
\begin{equation}
x^{'}_1 = a_1*x_1 + a_2*x_2 + a_3*x_3 \label{bigreg}
\end{equation}
\begin{equation}
x^{'}_1  = a_1*x_1 \mbox{,\quad} x^{'}_1 += a_2*x_2 \mbox{,\quad} x^{'}_1 += a_3*x_3 \label{smallreg}
\end{equation}

We can reduce the number of registers in use per thread down below 64
by using Eq.~\ref{smallreg}.
Usual matrix-vector multiplication is done by  Eq.~\ref{bigreg}. 
In that way, each thread needs total 77 registers.
But if we use Eq.~\ref{smallreg}, then, we need only 61
registers.
As a result, 2 threads blocks can be launched for each multiprocessor. 
This give us a 7\% gain.

Reducing the branch code\footnote{A branch code means part of the code
  which uses if, else if, switch, and so on.} also gives a better
performance.
GPU is not good in branch prediction.
Hence, removing unwanted branch code can increase the performance. 
As a result, it provides additional 30\% of performance enhancement to
our code.
In addition, we use bit operator and loop unrolling. 
Each of them gives a few \% of performance enhancements.
After all the optimizations, the final performance is about 35 giga flops, 
which is 47\% of the peak performance in GTX 295. 
The GPU code is 76 times faster than the CPU code at present.

\section{Multi GPU usage}
In our program, we use MPI for multi GPU implementation.
First, each node collects fermion vectors in lattice boundary on GPU
memory space.
Next, collected data are downloaded to host memory.
Then, data are transferred to neighbor nodes by infiniband network
(MPI).
Finally, data in the host memory is uploaded to GPU memory and repeat
the CG calculation.
Compared with single GPU case, using multi GPU needs network
communication by infiniband and PCI-E bus I/O.
Unfortunately, data transfer between GPU nodes makes the program slow
down dramatically as one can see in Table~\ref{time_stamp}.
\begin{table}[h]
\begin{center}
\begin{tabular}{|c|c|c|}
\hline
\quad & 4 node(ms) & 8 node(ms) \\
\hline
GPU calculation time      & 4.7 & 2.45 \\
boundary data collect     & 0.9 & 0.5 \\
gpu memory to host memory & 2.9 & 2.1 \\
MPI communication         & 2.3 & 1.8 \\
host memory to gpu memory & 3.4 & 1.7 \\
\hline
Total time & $\sim$ 14.5 & $\sim$ 8.6\\
\hline
\end{tabular}
\caption{Time stamps of CG program using multi GPUs.
Here, ms denotes mili seconds.}
\label{time_stamp}
\end{center}
\end{table}
In Table \ref{time_stamp}, we present elapsed time of each operation.
Here, the data transfer is dominant ($\approx$ 70\%). 
The floating point calculation of GPU occupies only 32\% in 4 node
calculation and the rest is for data transfer.
So we have to make the data transfer faster for better performance.
For better MPI performance, we use the MPI asynchronous network
communication mode so that the CPU can run the next jobs simultaneously
during the MPI data transfer .
Fortunately, CUDA also supports asynchronous data transfer between GPU
memory and CPU memory.
Hence, if we can overlap cudamemcpy with MPI communication by using
asynchronous communication, then total communication time can be
reduced by almost 1/3 in principle.

\section{Future Perspectives}
Unfortunately, the idea to overlap the network I/O with cudamemcpy I/O
is not working in our machine.
There are two problems.
First, simultaneous bi-directional memcpy between GPU and CPU is not
supported yet.
Our GPU is GTX 295 which does not support bi-directional memory copy
Therefore, GPU download sequence and upload sequence cannot be overlapped.
The second problem is that the page locked memory cannot be shared by
MPI and CUDA.
If we use asynchronous memcpy in CUDA or MPI simultaneously,
then each operation needs to use its own page locked memory.
But these memory regions cannot be shared by them yet. 
The two regions of page locked memory should be located in different
memory areas.
Hence, it needs additional memory copy step. This makes the
communication optimization more difficult.
But there are possibilities to solve this problem.
Actually, the 1st problem is not a problem anymore.
From the FERMI version of Tesla GPUs, they have 2 memory copy engines 
which support the bi-directional memory copy \cite{Fermi_white_paper}.
And 2nd problem can also be solved by using the GPU-direct technology
of Mellanox \cite{GPU_direct}.
GPU-direct technology enables us sharing the page locked
memory between GPU and MPI.
Hence, additional memory copy between two page locked areas is not
necessary.
Unfortunately, we have not tested this functions yet, because we don`t
have Mellanox device and nVIDIA Tesla GPU. We expect that our
machine (Qlogic) can also support these functions soon.

In near future, mixed precision method will be also implemented for
CG, which will enhance the performance significantly.

\section{Conclusion}
By using GPUs, we can get a good performance in the CG algorithm for
staggered fermions.
The final performance is about 35 giga flops per GPU (GTX 295). 
GPU is 75 times faster result then CPU.
We notice that data transfer between GPU and GPU memory is a
bottle-neck.
For better performance, various optimization methods are used. 
Including the MPI network communication, the performance is
reduced down to 12.3 giga flops in double precision.
At present, we continue working for the better performance.

The CUDA code of CG with multi GPUs is running in the production mode
to calculate hadron spectrum and weak matrix elements relevant to CP
violation in the neutral kaon system
\cite{ref:wlee-2010-1,ref:wlee-2010-2} at present.

\end{document}